\documentclass[proceedings]{stacs}
\stacsheading{2010}{597-608}{Nancy, France}
\firstpageno{597}

\usepackage[english]{babel}
\usepackage[latin1]{inputenc}
\usepackage{amsmath,amssymb,amsfonts,mathrsfs,latexsym,stmaryrd}
\usepackage{array}
\usepackage{url}
\usepackage{comment}
\usepackage[newitem,newenum,flushleft,neverdecrease]{paralist}
\usepackage{epsfig}
\usepackage{ifpdf}
\usepackage{color}
\definecolor{darkcyan}{rgb}{0,0.55,0.55}

\ifpdf
\else
\fi

\newenvironment{dotlist}{\begin{compactitem}}{\end{compactitem}}
\newenvironment{numlist}{\begin{compactenum}}{\end{compactenum}}
\newenvironment{romlist}{\begin{compactenum}[i)]}{\end{compactenum}}

\newcommand{\refcond}[1]{\hyperref[#1]{C\ref*{#1}}}
\newcommand{\refprop}[1]{\hyperref[#1]{P\ref*{#1}}}
\newcommand{\refaxiom}[1]{\hyperref[#1]{A\ref*{#1}}}

\providecommand{\qed}{\hfill $\Box$}
\renewcommand{\qed}{\hfill $\Box$}

\newcommand{\bbI}{\mathbb{I}}

\newcommand{\bbN}{\mathbb{N}}

\newcommand{\bbZ}{\mathbb{Z}}
\newcommand{\bbP}{\mathbb{P}}
\newcommand{\bbQ}{\mathbb{Q}}

\newcommand{\bbR}{\mathbb{R}}

\newcommand{\cG}{\mathcal{G}}

\newcommand{\cL}{\mathcal{L}}

\newcommand{\cO}{\mathcal{O}}
\newcommand{\cP}{\mathcal{P}}

\newcommand{\cR}{\mathcal{R}}
\newcommand{\cS}{\mathcal{S}}
\newcommand{\cT}{\mathcal{T}}

\newcommand{\sP}{\mathscr{P}}

\newcommand{\s}[1]{\vspace{#1mm}}

\providecommand{\mit}{\mathit}
\renewcommand{\mit}{\mathit}

\makeatletter
\def\shortrightarrowfill@{\arrowfill@\relbar\relbar\shortrightarrow}
\newcommand{\ort}{\mathpalette{\overarrow@\shortrightarrowfill@}}
\makeatother
\makeatletter
\def\shortleftarrowfill@{\arrowfill@\relbar\relbar\shortleftarrow}
\newcommand{\olft}{\mathpalette{\overarrow@\shortleftarrowfill@}}
\makeatother
\makeatletter
\def\shortleftrightarrowfill@{\arrowfill@\relbar\relbar\leftrightarrow}
\newcommand{\olftrt}{\mathpalette{\overarrow@\shortleftrightarrowfill@}}
\makeatother
\newcommand{\til}[1]{\widetilde{#1}}

\newcommand{\nin}{\not\in}
\newcommand{\sat}{\vDash}

\newcommand{\et}{\;\wedge\;}
\newcommand{\vel}{\;\vee\;}
\newcommand{\then}{\;\rightarrow\;}

\newcommand{\ang}[1]{{\langle #1 \rangle}}

\newcommand{\len}[1]{{\lvert #1 \rvert}}

\usepackage[OT1]{eulervm}

\renewcommand{\bar}[1]{\,\overline{\!#1\!}\,}
\newcommand{\DA}{\ang{A}}
\newcommand{\DB}{\ang{B}}

\newcommand{\DAbar}{\ang{\bar{A}}}
\newcommand{\DBbar}{\ang{\bar{B}}}

\newcommand{\BA}{[A]}
\newcommand{\BB}{[B]}

\newcommand{\BBbar}{[\bar{B}]}

\newcommand{\ABB}{A\mspace{-0.3mu}B\bar{B}}
\newcommand{\AEE}{\bar{A}\mspace{-0.3mu}E\bar{E}}
\newcommand{\AB}{A\mspace{-0.3mu}B}

\renewcommand{\prop}{\cP\mit{rop}}
\newcommand{\closure}{\mathcal{C}\mit{l}}
\newcommand{\eclosure}{\mathcal{C}\mit{l}^+}
\newcommand{\type}{\cT\mit{ype}}
\newcommand{\req}{\cR\mit{eq}}
\renewcommand{\obs}{\cO\mit{bs}}
\newcommand{\shading}{\cS\mit{hading}}
\newcommand{\labeledrightarrow}[1]{\overset{\text{\raisebox{-0.1ex}[0ex][-0.1ex]{$_{#1\,}$}}}{\longrightarrow}}

\newcommand{\dep}[1]{\,\text{\raisebox{-0.2ex}{$\labeledrightarrow{#1}$}}\,}

\begin{document}     
     
\title[Decidability of the interval temporal logic $\ABB$]{Decidability
 of the interval temporal \\ logic $\ABB$ over the natural numbers}
\author[inst1]{A. Montanari}{Angelo Montanari}
\address[inst1]{Department of Mathematics and Computer Science, Udine University, Italy}
\email{{angelo.montanari|pietro.sala}@dimi.uniud.it}
\author[inst2]{G. Puppis}{Gabriele Puppis}
\address[inst2]{Computing Laboratory, Oxford University, England}
\email{Gabriele.Puppis@comlab.ox.ac.uk}
\author[inst1]{P. Sala}{Pietro Sala}
\author[inst3]{G. Sciavicco}{Guido Sciavicco}
\address[inst3]{Department of Information, Engineering and Communications, Murcia University, Spain}
\email{guido@um.es}


\begin{abstract}
In this paper, we focus our attention on the interval temporal logic
of the Allen's relations ``meets'', ``begins'', and ``begun by'' ($\ABB$
for short), interpreted over natural numbers. We first introduce the logic and
we show that it is expressive enough to model distinctive interval properties,
such as accomplishment conditions, to capture basic modalities of point-based 
temporal logic, such as the until operator, and to encode relevant metric
constraints. Then, we prove that the satisfiability problem for $\ABB$ over 
natural numbers is decidable by providing a small model theorem based on an 
original contraction method. Finally, we prove the EXPSPACE-completeness 
of the problem\thanks{The work has been partially supported by the
GNCS project: ``Logics, automata, and games for the formal verification 
of complex systems''. Guido Sciavicco has also been supported by
the Spain/South Africa Integrated Action N.\ HS2008-0006 on: ``Metric 
interval temporal logics: Theory and Applications''.}.
\end{abstract}

\maketitle

\section{Introduction}\label{sec:intro}

Interval temporal logics are modal logics that allow one to represent and to reason 
about time intervals. It is well known that, on a linear ordering, one among thirteen 
different binary relations may hold between any pair of intervals, namely, ``ends'', 
``during'', ``begins'', ``overlaps'', ``meets'', ``before'', together with their 
inverses, and the relation ``equals'' (the so-called Allen's relations 
\cite{interval_relations})\footnote{We do not consider here the case of ternary 
                                    relations. Amongst the multitude of {\sl ternary} 
                                    relations among intervals there is one of particular 
                                    importance, which corresponds to the binary operation 
                                    of concatenation of meeting intervals. The logic of such 
                                    a ternary interval relation has been investigated by 
                                    Venema in \cite{chopping_intervals}. A systematic 
                                    analysis of its fragments has been recently given by 
                                    Hodkinson et al. \cite{cdt_axiomatizability}.}. 
Allen's relations give rise to respective unary modal operators, thus defining the 
modal logic of time intervals HS introduced by Halpern and Shoham in \cite{interval_modal_logic}. 
Some of these modal operators are actually definable in terms of others; in particular, if 
singleton intervals are included in the structure, it suffices to choose as basic the modalities 
corresponding to the relations ``begins'' $B$ and ``ends'' $E$, and their transposes $\bar{B}$, 
$\bar{E}$. HS turns out to be highly undecidable under very weak assumptions on the class of 
interval structures over which its formulas are interpreted \cite{interval_modal_logic}. In 
particular, undecidability holds for any class of interval structures over linear orderings 
that contains at least one linear ordering with an infinite ascending or descending chain, 
thus including the natural time flows $\bbN$, $\bbZ$, $\bbQ$, and $\bbR$. 
Undecidability of HS over finite structures directly follows from results 
in \cite{Lutz-Wolter-LMCS-06}. In \cite{undecidability_BE}, Lodaya sharpens the 
undecidability of HS showing that the two modalities $B, E$ suffice for 
undecidability over dense linear orderings (in fact, the result applies 
to the class of all linear orderings \cite{roadmap_intervals}).
Even though HS is very natural and the meaning of its operators is quite intuitive,
for a long time such sweeping undecidability results have 
discouraged the search for practical applications and further investigations in the field. A 
renewed interest in interval temporal logics has been recently stimulated by the identification 
of some decidable fragments of HS, whose decidability does not depend on simplifying semantic 
assumptions such as locality and homogeneity \cite{roadmap_intervals}. This is the case with the 
fragments $B\bar{B}$, $E\bar{E}$ (logics of the ``begins/begun by'' and ``ends/ended by'' relations) 
\cite{roadmap_intervals}, $A$, $A\bar{A}$ (logics of temporal neighborhood, whose modalities
capture the ``meets/met by'' relations \cite{pnl_logics}), and 
$D$, $D\bar{D}$ (logics of the subinterval/superinterval relations) \cite{subinterval_tableau_journal,cone_logic}. 

In this paper, we focus our attention on the product logic $\ABB$, obtained from the join of 
$B\bar{B}$ and $A$ (the case of $\AEE$ is fully symmetric), interpreted over the linear order
$\bbN$ of the natural numbers (or a finite prefix of it).
The decidability of $B\bar{B}$ can be proved by translating it into the point-based propositional  
temporal logic of linear time with temporal modalities $F$ (sometime in the future) and $P$ 
(sometime in the past), which has the finite (pseudo-)model property and is decidable, e.g., 
\cite{temporal_logic_foundations}.
In general, such a reduction to point-based temporal logics does not work: formulas of  interval 
temporal logics  are evaluated over pairs of points and translate into binary relations. For instance, 
this is the case with $A$. Unlike the case of $B\bar{B}$, when dealing with $A$ one cannot abstract way 
from the left endpoint of intervals, as contradictory formulas may hold over intervals with the 
same right endpoint and a different left endpoint.
The decidability of $A\bar{A}$, and thus that of its fragment $A$, over various classes of linear 
orderings has been proved by Bresolin et al. by reducing its satisfiability problem to that of 
the two-variable fragment of first-order logic over the same classes of structures 
\cite{pnl_decidability_paper}, whose decidability has been proved by Otto in \cite{two_variable_fo}. 
Optimal tableau methods for $A$ with respect to various classes of interval structures can be 
found in \cite{optimal_rpnl_tableau,tableau_for_right_pnl}. A decidable metric extension of $A$
over the natural numbers has been proposed in \cite{sefm2009}. A number of undecidable extensions 
of $A$, and $A\bar{A}$, have been given in \cite{halpern_shoham_fragments,pnl_expressiveness}.

$\ABB$ retains the simplicity of its constituents $B\bar{B}$ and $A$, but it improves a lot on 
their expressive power (as we shall show, such an increase in expressiveness is achieved at the 
cost of an increase in complexity). First, it allows one to express assertions that may be true 
at certain intervals, but at no subinterval of them, such as the conditions of accomplishment. 
Moreover, it makes it possible to easily encode the until operator of point-based temporal logic 
(this is possible neither with $B\bar{B}$ nor with $A$). Finally, meaningful metric constraints 
about the length of intervals can be expressed in $\ABB$, that is, one can constrain an interval 
to be at least (resp., at most, exactly) $k$ points long. 
We prove the decidability of $\ABB$ interpreted over $\bbN$ by providing a small model theorem based 
on an original contraction method. To prove it, we take advantage of a natural (equivalent) 
interpretation of $\ABB$ formulas over grid-like structures based on a bijection between the set 
of intervals over $\bbN$ and (a suitable subset of) the set of points of the $\bbN \times \bbN$ grid. In 
addition, we prove that the satisfiability problem for $\ABB$ is EXPSPACE-complete (that for $A$ 
is NEXPTIME-complete). In the proof of hardness, we use a reduction from the exponential-corridor 
tiling problem.

The paper is organized as follows. In Section \ref{sec:logic} we introduce $\ABB$. In Section 
\ref{sec:smallmodel_abb}, we prove the decidability of its satisfiability problem. We first 
describe the application of the contraction method to finite models and then we generalize
it to infinite ones. In Section \ref{sec:completeness} we deal with computational complexity 
issues. Conclusions provide an assessment of the work and outline future research directions. 
Missing proofs can be found in \cite{abb_report}.

\section{The interval temporal logic $\ABB$}\label{sec:logic}

In this section, we briefly introduce syntax and semantics of the logic $\ABB$, which features 
three modal operators $\DA$, $\DB$, and $\DBbar$ corresponding to the three Allen's relations  
$A$ (``meets''), $B$ (``begins''), and $\bar{B}$ (``begun by''), respectively. We show that $\ABB$ is 
expressive enough to capture the notion of accomplishment, to define the standard until operator 
of point-based temporal logics, and to encode metric conditions. Then, we introduce the basic 
notions of atom, type, and dependency. We conclude the section by providing an alternative 
interpretation of $\ABB$ over labeled grid-like structures.

\subsection{Syntax and semantics}\label{subsec:syntax}

Given a set $\prop$ of propositional variables, formulas of $\ABB$ are built up from $\prop$ using 
the boolean connectives $\neg$ and $\vel$ and the unary modal operators $\DA$, $\DB$, $\DBbar$. As 
usual, we shall take advantage of shorthands like $\varphi_1\et\varphi_2=\neg(\neg\varphi_1 \vel 
\neg\varphi_2)$, $\BA\varphi=\neg\DA\neg\varphi$, $\BB\varphi=\neg\DB\neg\varphi$, $\top = p \vee \neg p$,
and $\bot = p \wedge \neg p$, with $p \in \prop$. Hereafter, we denote by $\len{\varphi}$ the size of $\varphi$.

We interpret formulas of $\ABB$ in interval temporal structures over natural numbers endowed with
the relations ``meets'', ``begins'', and ``begun by''. Precisely, we identify any given ordinal 
$N\le\omega$ with the prefix of length $N$ of the linear order of the natural numbers and we 
accordingly define $\bbI_N$ as the set of all non-singleton closed intervals $[x,y]$, with $x,y\in N$ 
and $x<y$. For any pair of intervals $[x,y], [x',y'] \in \bbI_N$, the Allen's relations ``meets'' 
$A$, ``begins'' $B$, and ``begun by'' $\bar{B}$ are defined as follows (note that $\bar{B}$ is 
the inverse relation of $B$):
\begin{dotlist}
  \item {\bf ``meets'' relation:} $[x,y] \;A\; [x',y']$ iff $y=x'$;
  \item {\bf ``begins'' relation:} $[x,y] \;B\; [x',y']$ iff $x=x'$ and $y'<y$;
  \item {\bf ``begun by'' relation:} $[x,y] \;\bar{B}\; [x',y']$ iff $x=x'$ and $y<y'$.
\end{dotlist}
Given an \emph{interval structure} $\cS=(\bbI_N,A,B,\bar{B},\sigma)$, where $\sigma:\bbI_N\then\sP(\prop)$ 
is a labeling function that maps intervals in $\bbI_N$ to sets of propositional variables, and an initial 
interval $I$, we define the semantics of an $\ABB$ formula as follows:
\begin{dotlist}
  \item $\cS,I \sat a$ iff $a\in\sigma(I)$, for any $a\in\prop$;
  \item $\cS,I \sat \neg\varphi$ iff $\cS,I \not\sat\varphi$;  
  \item $\cS,I \sat \varphi_1 \vel \varphi_2$ iff $\cS,I \sat \varphi_1$ or $\cS,I \sat \varphi_2$;
  \item for every relation $R\in\{A,B,\bar{B}\}$, $\cS,I \sat \ang{R}\varphi$ iff there is an interval 
        $J\in\bbI_N$ such that $I \;R\; J$ and $\cS,J \sat \varphi$.
\end{dotlist}
Given an interval structure $\cS$ and a formula $\varphi$, we say that $\cS$ \emph{satisfies} $\varphi$ 
if there is an interval $I$ in $\cS$ such that $\cS,I\sat\varphi$. We say that $\varphi$ is \emph{satisfiable}
if there exists an interval structure that satisfies it. We define the \emph{satisfiability problem} for 
$\ABB$ as the problem of establishing whether a given $\ABB$-formula $\varphi$ is satisfiable.

We conclude the section with some examples that account for $\ABB$ expressive power. The first one shows how 
to encode in $\ABB$ conditions of accomplishment (think of formula $\varphi$ as the assertion: ``Mr. Jones 
flew from Venice to Nancy''): $\DA\bigl(\varphi \et \BB (\neg \varphi \et \BA \neg \varphi) \et \BBbar \neg \varphi \bigr)$. 
Formulas of point-based temporal logics of the form $\psi \;U\; \varphi$, using the standard until operator, 
can be encoded in $\ABB$ (where atomic intervals are two-point intervals) as follows: 
$\DA\bigl(\BB\bot \!\et\! \varphi\bigr) \vel \DA\bigl(\DA(\BB\bot \!\et\! \varphi) \et \BB(\DA(\BB\bot \!\et\! \psi))\bigr).$
Finally, metric conditions like: ``$\varphi$ holds over a right neighbor interval of length greater than $k$ (resp., 
less than $k$, equal to $k$)'' can be captured by the following $\ABB$ formula: $\DA\bigl(\varphi \et \DB^k\top\bigr)$ 
(resp., $\DA\bigl(\varphi \et \BB^{k-1}\bot\bigr)$, $\DA\bigl(\varphi \et \BB^k\bot \et 
\DB^{k-1}\top\bigr)$)\footnote{It is not difficult to show that $\ABB$ subsumes the metric
extension of $A$ given in \cite{sefm2009}. A simple game-theoretic argument shows that the former is 
in fact strictly more expressive than the latter.}.

\subsection{Atoms, types, and dependencies}\label{subsec:types}

Let $\cS=(\bbI_N,A,B,\bar{B},\sigma)$ be an interval structure and $\varphi$ be a formula of $\ABB$.
In the sequel, we shall compare intervals in $\cS$ with respect to the set of subformulas of $\varphi$ 
they satisfy. To do that, we introduce the key notions of $\varphi$-atom, $\varphi$-type, $\varphi$-cluster, 
and $\varphi$-shading. 

First of all, we define the \emph{closure} $\closure(\varphi)$ of $\varphi$ as the set of all 
subformulas of $\varphi$ and of their negations (we identify $\neg\neg\alpha$ with $\alpha$, 
$\neg\DA\alpha$ with $\BA\neg\alpha$, etc.). For technical reasons, we also introduce the 
\emph{extended closure} $\eclosure(\varphi)$, which is defined as the set of all formulas 
in $\closure(\varphi)$ plus all formulas of the forms $\ang{R}\alpha$ and $\neg\ang{R}\alpha$, 
with $R\in\{A,B,\bar{B}\}$ and $\alpha\in\closure(\varphi)$.

A \emph{$\varphi$-atom} is any non-empty set $F\subseteq\eclosure(\varphi)$ such that (i) for every 
$\alpha\in\eclosure(\varphi)$, we have $\alpha\in F$ iff $\neg\alpha\nin F$ and (ii) for every 
$\gamma=\alpha\vel\beta\in\eclosure(\varphi)$, we have $\gamma\in F$ iff $\alpha\in F$ or $\beta\in F$
(intuitively, a \emph{$\varphi$-atom} is a maximal {\sl locally consistent} set of formulas chosen
from $\eclosure(\varphi)$). Note that the cardinalities of both sets $\closure(\varphi)$ and 
$\eclosure(\varphi)$ are {\sl linear} in the number $\len{\varphi}$ of subformulas of $\varphi$, 
while the number of $\varphi$-atoms is {\sl at most exponential} in $\len{\varphi}$ (precisely, 
we have $\len{\closure(\varphi)}=2\len{\varphi}$, $\len{\eclosure(\varphi)}=14\len{\varphi}$, and
there are at most $2^{7\len{\varphi}}$ distinct atoms).

We also associate with each interval $I\in\cS$ the set of all formulas $\alpha\in\eclosure(\varphi)$ 
such that $\cS,I\sat\alpha$. Such a set is called \emph{$\varphi$-type} of $I$ and it is denoted by 
$\type_\cS(I)$. We have that every $\varphi$-type is a $\varphi$-atom, but not vice versa. Hereafter, 
we shall omit the argument $\varphi$, thus calling a $\varphi$-atom (resp., a $\varphi$-type) simply 
an atom (resp., a type).

Given an atom $F$, we denote by $\obs(F)$ the set of all \emph{observables} of $F$, namely, the 
formulas $\alpha\in\closure(\varphi)$ such that $\alpha\in F$. Similarly, given an atom $F$ and 
a relation $R\in\{A,B,\bar{B}\}$, we denote by $\req_R(F)$ the set of all \emph{$R$-requests} 
of $F$, namely, the formulas $\alpha\in\closure(\varphi)$ such that $\ang{R}\alpha\in F$. Taking 
advantage of the above sets, we can define the following two relations between atoms $F$ 
and $G$:
$$
\begin{array}{rcl}
  F \dep{A} G    
  &\;\quad\text{iff}\quad\quad&   
  \req_A(F)                              \;=\;          \obs(G) \,\cup\, \req_B(G) \,\cup\, \req_{\bar{B}}(G);    \s2 \\
  F \dep{B} G    
  &\;\quad\text{iff}\quad\quad&   
  \begin{cases}   
    \obs(F) \,\cup\, \req_{\bar{B}}(F)   \;\subseteq\;  \req_{\bar{B}}(G)                                      
                                         \;\subseteq\;  \obs(F) \,\cup\, \req_{\bar{B}}(F) \,\cup\, \req_B(F),    \s1 \\
    \obs(G) \,\cup\, \req_B(G)           \;\subseteq\;  \req_B(F)
                                         \;\subseteq\;  \obs(G) \,\cup\, \req_B(G) \,\cup\, \req_{\bar{B}}(G).  
  \end{cases}
\end{array}
$$
Note that the relation $\dep{B}{}$ is transitive, while $\dep{A}{}$ is not. Moreover, both $\dep{A}{}$ 
and $\dep{B}{}$ satisfy a \emph{view-to-type dependency}, namely, for every pair of intervals $I,J$ in 
$\cS$, we have that
$$
\begin{array}{rcl}
  I \;A\; J    &\;\quad\text{implies}\quad\quad&   \type_\cS(I) \,\dep{A}{}\, \type_\cS(J)     \s1 \\
  I \;B\; J    &\;\quad\text{implies}\quad\quad&   \type_\cS(I) \,\dep{B}{}\, \type_\cS(J).
\end{array}
$$
Relations $\dep{A}{}$ and $\dep{B}{}$ will come into play in the definition of 
consistency conditions (see Definition \ref{def:compassstructure}).

\subsection{Compass structures}\label{subsec:compass}

\begin{figure}[!!t]
\centering
\includegraphics[scale=0.95]{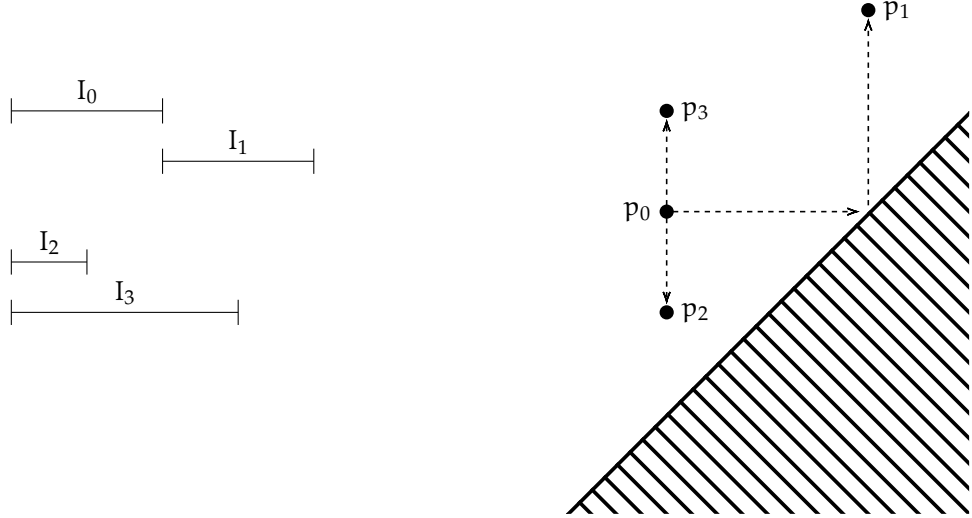}
\caption{Correspondence between intervals and points of a discrete grid.}
\label{fig:compassstructure}
\end{figure}
The logic $\ABB$ can be equivalently interpreted over grid-like 
structures (the so-called compass structures \cite{chopping_intervals})
by exploiting the existence of a natural bijection between the intervals $I=[x,y]$ and the points $p=(x,y)$ of 
an $N\times N$ grid such that $x<y$. As an example, Figure \ref{fig:compassstructure} depicts four intervals
$I_0,...,I_3$ such that $I_0 \;A\; I_1$, $I_0 \;B\; I_2$, and $I_0 \;\bar{B}\;I_3$, together with the corresponding 
points $p_0,...,p_3$ of a discrete grid (note that the three Allen's relations $A,B,\bar{B}$ between intervals 
are mapped to corresponding spatial relations between points; for the sake of readability, we name the latter 
ones as the former ones).

\begin{definition}\label{def:compassstructure}
Given an $\ABB$ formula $\varphi$, a (consistent and fulfilling) \emph{compass} ($\varphi$-)\emph{structure} 
of length $N\le\omega$ is a pair $\cG=(\bbP_N,\cL)$, where $\bbP_N$ is the set of points $p=(x,y)$, with 
$0\le x<y<N$, and $\cL$ is function that maps any point $p\in\bbP_N$ to a ($\varphi$-)atom $\cL(p)$ in such a way that 
\begin{dotlist}
  \item for every pair of points $p,q\in\bbP_N$ and every relation $R\in\{A,B\}$,
        if $p \;R\; q$ holds, then $\cL(p) \dep{R} \cL(q)$ follows ({\bf consistency});
  \item for every point $p\in\bbP_N$, every relation $R\in\{A,B,\bar{B}\}$, and 
        every formula $\alpha\in\req_R\bigl(\cL(p)\bigr)$, there is a point $q\in\bbP_N$ such that 
        $p \;R\; q$ and $\alpha\in\obs\bigl(\cL(q)\bigr)$ ({\bf fulfillment}).
\end{dotlist}
\end{definition}

\noindent
We say that a compass ($\varphi$-)structure $\cG=(\bbP_N,\cL)$ \emph{features} a formula
$\alpha$ if there is a point $p\in\bbP_N$ such that $\alpha \in \cL(p)$. The following 
proposition implies that the satisfiability problem for $\ABB$ is reducible to the problem 
of deciding, for any given formula $\varphi$, whether there exists a $\varphi$-compass 
structure that features $\varphi$. 

\begin{proposition}\label{prop:compassstructure}
An $\ABB$-formula $\varphi$ is satisfied by some interval structure if and only if it is 
featured by some ($\varphi$-)compass structure.
\end{proposition}

\section{Deciding the satisfiability problem 
for $\ABB$}
\label{sec:smallmodel_abb}

In this section, we prove that the satisfiability problem for $\ABB$ is decidable 
by providing a ``small-model theorem'' for the satisfiable formulas of the logic. 
For the sake of simplicity, we first show that the satisfiability problem for
$\ABB$ interpreted over {\sl finite} interval structures is decidable and then we 
generalize such a result to all (finite or infinite) interval structures.

As a preliminary step, we introduce the key notion of shading. Let $\cG=(\bbP_N,\cL)$ be a compass 
structure of length $N\le\omega$ and let $0\le y<N$. The \emph{shading of the row $y$ of $\cG$}
is the set $\shading_\cG(y)=\bigl\{\cL(x,y) \,:\, 0\le x<y\bigr\}$, namely, the set of the atoms 
of all points in $\bbP_N$ whose vertical coordinate has value $y$ (basically, we interpret 
different atoms as different colors). Clearly, for every pair of atoms $F$ and $F'$ in 
$\shading_\cG(y)$, we have $\req_A(F)=\req_A(F')$.

\subsection{A small-model theorem for finite structures}\label{subsec:finitecase}

Let $\varphi$ be an $\ABB$ formula. Let us assume that $\varphi$ is featured by a 
{\sl finite} compass structure $\cG=(\bbP_N,\cL)$, with $N<\omega$. In fact, without loss 
of generality, we can assume that $\varphi$ belongs to the atom associated with a point 
$p=(0,y)$ of $\cG$, with $0<y<N$. We prove that we can restrict our attention to compass 
structures $\cG=(\bbP_N,\cL)$, where $N$ is bounded by a double exponential in $\len{\varphi}$.
We start with the following lemma that proves a simple, but crucial, property of the relations 
$\dep{A}$ and $\dep{B}$ (the proof can be found in \cite{abb_report}).

\begin{lemma}\label{lemma:entanglement}
If $F\dep{A}H$ and $G\dep{B}H$ hold for some atoms $F,G,H$, then $F\dep{A}G$ holds as well.
\end{lemma}

\medskip
The next lemma shows that, under suitable conditions, a given compass structure
$\cG$ may be reduced in length, preserving the existence of atoms featuring $\varphi$.

\begin{lemma}\label{lemma:contraction_finite}
Let $\cG$ be a compass structure featuring $\varphi$. If there exist two rows $0<y_0<y_1<N$ 
in $\cG$ such that $\shading_\cG(y_0)\subseteq\shading_\cG(y_1)$, then there exists a compass 
structure $\cG'$ of length $N'<N$ that features $\varphi$.
\end{lemma}

\begin{figure}[!!t]
\centering
\includegraphics[scale=0.9]{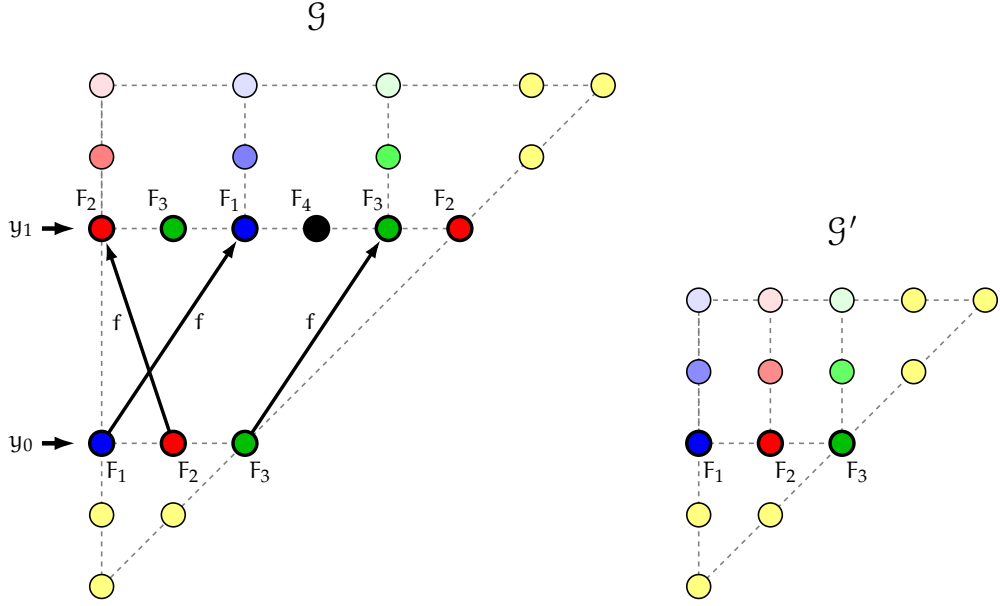}
\caption{Contraction $\cG'$ of a compass structure $\cG$.}
\label{fig:contraction}
\end{figure}
\begin{proof}
Suppose that $0<y_0<y_1<N$ are two rows of $\cG$ such that $\shading_\cG(y_0)\subseteq\shading_\cG(y_1)$.
Then, there is a function $f:\{0,...,y_0-1\}\then\{0,...,y_1-1\}$ such that,
for every $0\le x<y_0$, $\cL(x,y_0)=\cL(f(x),y_1)$. Let $k=y_1-y_0$, $N'=N-k$ ($<N$), and 
$\bbP_{N'}$ be the portion of the grid that consists of all points $p=(x,y)$, with $0\le x<y<N'$.
We extend $f$ to a function that maps points in $\bbP_{N'}$ to points in $\bbP_N$ as follows:
\begin{dotlist}
  \item if $p=(x,y)$, with $0\le x<y<y_0$, then we simply let $f(p)=p$;
  \item if $p=(x,y)$, with $0\le x<y_0\le y$, then we let $f(p)=(f(x),y+k)$;
  \item if $p=(x,y)$, with $y_0\le x<y$, then we let $f(p)=(x+k,y+k)$.
\end{dotlist}
We denote by $\cL'$ the labeling of $\bbP_{N'}$ such that, for every point $p\in\bbP_{N'}$,
$\cL'(p)=\cL(f(p))$ and we denote by $\cG'$ the resulting structure $(\bbP_{N'},\cL')$ (see Figure
\ref{fig:contraction}). We have to prove that $\cG'$ is a {\sl consistent} and {\sl fulfilling}
compass structure that features $\varphi$ (see Definition \ref{def:compassstructure}). First, we 
show that $\cG'$ satisfies the consistency conditions for the relations $B$ and $A$; then we show 
that $\cG'$ satisfies the fulfillment conditions for the $\bar{B}$-, $B$-, and $A$-requests; 
finally, we show that $\cG'$ features $\varphi$.

\smallskip\noindent
\textsc{Consistency with relation $B$.\;\;} 
Consider two points $p=(x,y)$ and $p'=(x',y')$ in $\cG'$ such that $p \;B\; p'$, i.e., $0\le x=x'<y'<y<N'$. 
We prove that $\cL'(p) \dep{B} \cL'(p')$ by distinguishing among the following three cases (note that 
exactly one of such cases holds): 
\begin{numlist}
  \item $y<y_0$ and $y'<y_0$,
  \item $y\ge y_0$ and $y'\ge y_0$, 
  \item $y\ge y_0$ and $y'<y_0$. 
\end{numlist}
If $y<y_0$ and $y'<y_0$, then, by construction, we have $f(p)=p$ and $f(p')=p'$. Since $\cG$ 
is a (consistent) compass structure, we immediately obtain $\cL'(p)=\cL(p) \dep{B} \cL(p')=\cL'(p')$.

\noindent
If $y\ge y_0$ and $y\ge y_0$, then, by construction, we have either $f(p)=(f(x),y+k)$ or $f(p)=(x+k,y+k)$, 
depending on whether $x<y_0$ or $x\ge y_0$. Similarly, we have either $f(p')=(f(x'),y'+k)=(f(x),y'+k)$ or 
$f(p')=(x'+k,y'+k)=(x+k,y'+k)$. This implies $f(p) \;B\; f(p')$ and thus, since $\cG$ is a (consistent) 
compass structure, we have $\cL'(p)=\cL(f(p))$ $\dep{B}$ $\cL(f(p'))=\cL'(p')$.

\noindent
If $y\ge y_0$ and $y'<y_0$, then, since $x<y'<y_0$, we have by construction $f(p)=(f(x),y+k)$ and $f(p')=p'$. 
Moreover, if we consider the point $p''=(x,y_0)$ in $\cG'$, we easily see that (i) $f(p'')=(f(x),y_1)$, 
(ii) $f(p) \;B\; f(p'')$ (whence $\cL(f(p)) \dep{B} \cL(f(p''))$), (iii) $\cL(f(p''))=\cL(p'')$, and 
(iv) $p'' \;B\; p'$ (whence $\cL(p'') \dep{B} \cL(p')$). It thus follows that 
$\cL'(p)=\cL(f(p)) \dep{B} \cL(f(p''))$ $=\cL(p'')$ $\dep{B}$ $\cL(p')=\cL(f(p'))=\cL'(p')$. Finally, 
by exploiting the transitivity of the relation $\dep{B}$, we obtain $\cL'(p) \dep{B} \cL'(p')$.

\smallskip\noindent
\textsc{Consistency with relation $A$.\;\;} 
Consider two points $p=(x,y)$ and $p'=(x',y')$ such that $p \;A\; p'$, i.e., $0\le x<y=x'<y'<N'$. 
We define $p''=(y,y+1)$ in such a way that $p \;A\; p''$ and $p' \;B\; p''$ and we distinguish between 
the following two cases:
\begin{numlist}
  \item $y\ge y_0$,
  \item $y<y_0$.
\end{numlist}
If $y\ge y_0$, then, by construction, we have $f(p) \;A\; f(p'')$. Since $\cG$ is a (consistent)
compass structure, it follows that $\cL'(p)=\cL(f(p))$ $\dep{A}$ $\cL(f(p''))=\cL'(p'')$. 

\noindent
If $y<y_0$, 
then, by construction, we have $\cL(p'')=\cL(f(p''))$. Again, since $\cG$ is a (consistent) compass 
structure, it follows that $\cL'(p)=\cL(f(p))=\cL(p)$ $\dep{A}$ $\cL(p'')=\cL(f(p''))=\cL'(p'')$.

\noindent
In both cases we have $\cL'(p) \dep{A} \cL'(p'')$. Now, we recall that $p' \;B\; p''$ and that, by 
previous arguments, $\cG'$ is consistent with the relation $B$. We thus have $\cL'(p') \dep{B} \cL'(p'')$. 
Finally, by applying Lemma \ref{lemma:entanglement}, we obtain $\cL'(p) \dep{A} \cL'(p')$.

\smallskip\noindent
\textsc{Fulfillment of $B$-requests.\;\;}
Consider a point $p=(x,y)$ in $\cG'$ and some $B$-request $\alpha\in\req_B\bigl(\cL'(p)\bigr)$ associated 
with it. Since, by construction, $\alpha\in\req_B\bigl(\cL(f(p))\bigr)$ and $\cG$ is a (fulfilling) 
compass structure, we know that $\cG$ contains a point $q'=(x',y')$ such that $f(p) \;B\; q'$ and 
$\alpha\in\obs\bigl(\cL(q')\bigr)$. We prove that $\cG'$ contains a point $p'$ such that $p \;B\; p'$ 
and $\alpha\in\obs\bigl(\cL'(p')\bigr)$ by distinguishing among the following three cases (note that 
exactly one of such cases holds):
\begin{numlist}
  \item $y<y_0$
  \item $y'\ge y_1$,
  \item $y\ge y_0$ and $y'<y_1$.
\end{numlist}
If $y<y_0$, then, by construction, we have $p=f(p)$ and $q'=f(q')$. Therefore, we simply 
define $p'=q'$ in such a way that $p=f(p) \;B\; q'=p'$ and $\alpha\in\obs\bigl(\cL'(p')\bigr)$ 
($=\obs\bigl(\cL(f(p'))\bigr)=\obs\bigl(\cL(q')\bigr)$).

\noindent
If $y'\ge y_1$, then, by construction, we have either $f(p)=(f(x),y+k)$ or $f(p)=(x+k,y+k)$, 
depending on whether $x<y_0$ or $x\ge y_0$. We define $p'=(x,y'-k)$ in such a way that $p \;B\; p'$. 
Moreover, we observe that either $f(p')=(f(x),y')$ or $f(p')=(x+k,y')$, depending on whether $x<y_0$ 
or $x\ge y_0$, and in both cases $f(p')=q'$ follows. This shows that $\alpha\in\obs\bigl(\cL'(p')\bigr)$ 
($=\obs\bigl(\cL(f(p')\bigr)=\obs\bigl(\cL(q')\bigr)$).

\noindent
If $y\ge y_0$ and $y'<y_1$, then we define $\bar{p}=(x,y_0)$ and $\bar{q}=(x',y_1)$ and we observe
that $f(p) \;B\; \bar{q}$, $\bar{q} \;B\; q'$, and $f(\bar{p})=\bar{q}$. From $f(p) \;B\; \bar{q}$
and $\bar{q} \;B\; q'$, it follows that $\alpha\in\req_B\bigl(\cL(\bar{q})\bigr)$ and hence 
$\alpha\in\req_B\bigl(\cL(\bar{p})\bigr)$. 
Since $\cG$ is a (fulfilling) compass structure, we know that there is a point $p'$ such that 
$\bar{p} \;B\; p'$ and $\alpha\in\obs\bigl(\cL(\bar{p}')\bigr)$. Moreover, since $\bar{p} \;B\; p'$, 
we have $f(p')=p'$, from which we obtain $p \;B\; p'$ and $\alpha\in\obs\bigl(\cL(p')\bigr)$.

\smallskip\noindent
\textsc{Fulfillment of $\bar{B}$-requests.\;\;}
The proof that $\cG'$ fulfills all $\bar{B}$-requests of its atoms is symmetric with respect to 
the previous one.

\smallskip\noindent
\textsc{Fulfillment of $A$-requests.\;\;}
Consider a point $p=(x,y)$ in $\cG'$ and some $A$-request $\alpha\in\req_A\bigl(\cL'(p)\bigr)$ 
associated with $p$ in $\cG'$. Since, by previous arguments, $\cG'$ fulfills all $\bar{B}$-requests
of its atoms, it is sufficient to prove that either $\alpha\in\obs\bigl(\cL'(p')\bigr)$ or 
$\alpha\in\req_{\bar{B}}\bigl(\cL'(p')\bigr)$, where $p'=(y,y+1)$. This can be easily proved by
distinguishing among the three cases $y<y_0-1$, $y=y_0-1$, and $y\ge y_0$.

\smallskip\noindent
\textsc{Featured formulas.\;\;}
Recall that, by previous assumptions, $\cG$ contains a point $p=(0,y)$, with $0<y<N$, such that
$\varphi\in\cL(p)$. If $y\le y_0$, then, by construction, we have $\varphi\in\cL'(p)$ ($=\cL(f(p))=\cL(p)$).
Otherwise, if $y>y_0$, we define $q=(0,y_0)$ and we observe that $q \;\bar{B}\; p$. Since $\cG$ is a 
(consistent) compass structure and $\ang{\bar{B}}\varphi\in\eclosure(\varphi)$, we have that
$\varphi\in\req_{\bar{B}}\bigl(\cL(q)\bigr)$. Moreover, by construction, we have $\cL'(q)=\cL(f(q))$
and hence $\varphi\in\req_{\bar{B}}\bigl(\cL'(q)\bigr)$. Finally, since $\cG'$ is a (fulfilling)
compass structure, we know that there is a point $p'$ in $\cG'$ such that $f(q) \;\bar{B}\; p'$ 
and $\varphi\in\obs\bigl(\cL'(p')\bigr)$.
\end{proof}

\medskip
On the grounds of the above result, we can provide a suitable upper bound for the length 
of a minimal finite interval structure that satisfies $\varphi$, if there exists any.
This yields a straightforward, but inefficient, 2EXPSPACE algorithm that decides 
whether a given $\ABB$-formula $\varphi$ is satisfiable over finite interval structures.

\begin{theorem}\label{th:contraction_finite}
An $\ABB$-formula $\varphi$ is satisfied by some finite interval structure iff it is 
featured by some compass structure of length $N\le 2^{2^{7\len{\varphi}}}$ (i.e., double 
exponential in $\len{\varphi}$).
\end{theorem}

\begin{proof}
One direction is trivial. We prove the other one (``only if'' part). Suppose that $\varphi$ is 
satisfied by a finite interval structure $\cS$. By Proposition \ref{prop:compassstructure}, there 
is a compass structure $\cG$ that features $\varphi$ and has finite length $N<\omega$. Without loss 
of generality, we can assume that $N$ is minimal among all finite compass structures that feature $\varphi$. 
We recall from Section \ref{subsec:types} that $\cG$ contains at most $2^{7\len{\varphi}}$ distinct 
atoms. This implies that there exist at most $2^{2^{7\len{\varphi}}}$ different shadings of the form 
$\shading_\cG(y)$, with $0\le y<N$. Finally, by applying Lemma \ref{lemma:contraction_finite}, 
we obtain $N\le 2^{2^{7\len{\varphi}}}$ (otherwise, there would exist two rows $0<y_0<y_1<N$ 
such that $\shading_\cG(y_0)=\shading_\cG(y_1)$, which is against the hypothesis of minimality 
of $N$).
\end{proof}

\subsection{A small-model theorem for infinite structures}\label{subsec:infinitecase}

In general, compass structures that feature $\varphi$ may be infinite. Here, we prove
that, without loss of generality, we can restrict our attention to sufficiently ``regular''
infinite compass structures, which can be represented in double exponential space with respect
to $\len{\varphi}$. To do that, we introduce the notion of periodic compass structure.

\begin{definition}\label{def:periodiccompassstructure}
An infinite compass structure $\cG=(\bbP_\omega,\cL)$ is \emph{periodic}, 
with \emph{threshold} $\til{y}_0$, \emph{period} $\til{y}$, and \emph{binding} 
$\til{g}:\{0,...,\til{y}_0+\til{y}-1\}\then\{0,...,\til{y}_0-1\}$, if the following
conditions are satisfied:
\begin{dotlist}
  \item for every $\til{y}_0+\til{y}\le x<y$, we have $\cL(x,y)=\cL(x-\til{y},y-\til{y})$,
  \item for every $0\le x<\til{y}_0+\til{y}\le y$, we have $\cL(x,y)=\cL(\til{g}(x),y-\til{y})$.
\end{dotlist}
\end{definition}

\begin{figure}[!!t]
\centering
\includegraphics[scale=0.95]{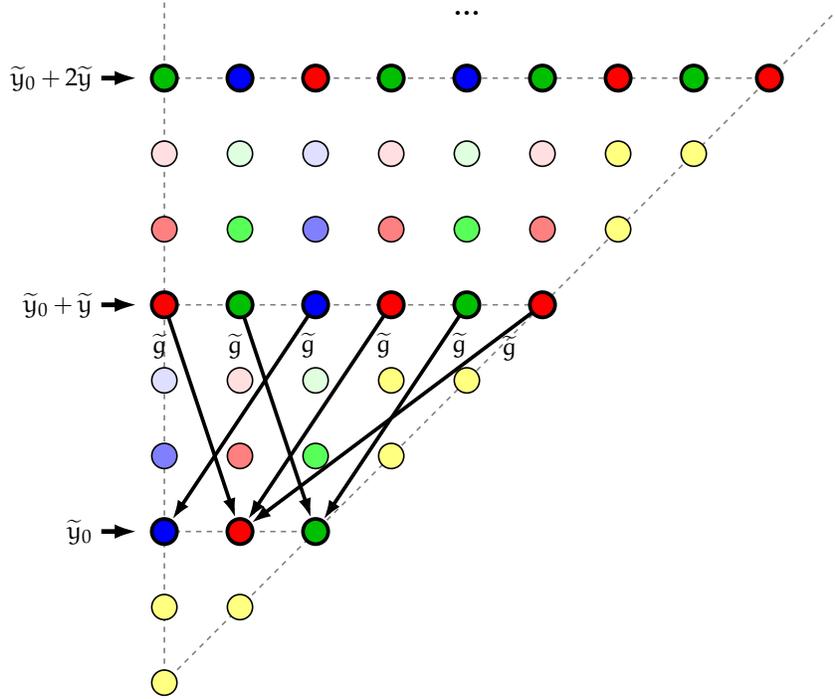}
\caption{A periodic compass structure with threshold $\til{y}_0$, period $\til{y}$, and binding $\til{g}$.}
\label{fig:periodiccompassstructure}
\end{figure}

\noindent
Figure \ref{fig:periodiccompassstructure} gives an example of a periodic compass structure (the 
arrows represent some relationships between points induced by the binding function $\til{g}$).
Note that any periodic compass structure $\cG=(\bbP_\omega,\cL)$ can be finitely represented by 
specifying (i) its threshold $\til{y}_0$, (ii) its period $\til{y}$, (iii) its binding $\til{g}$, 
and (iv) the labeling $\cL$ restricted to the portion $\bbP_{\til{y}_0+\til{y}-1}$ of the domain.

The following theorem leads immediately to a 2EXPSPACE algorithm that decides whether a given 
$\ABB$-formula $\varphi$ is satisfiable over infinite interval structures (the proof is provided 
in \cite{abb_report}).

\begin{theorem}\label{th:contraction_infinite}
An $\ABB$-formula $\varphi$ is satisfied by an infinite interval structure iff it is featured 
by a periodic compass structure with threshold $\til{y}_0<2^{2^{7\len{\varphi}}}$ and period 
$\til{y}<2\len{\varphi}\cdot 2^{2^{7\len{\varphi}}}\cdot 2^{2^{7\len{\varphi}}}$. 
\end{theorem}

\section{Tight complexity bounds to the satisfiability problem for $\ABB$}\label{sec:completeness}

In this section, we show that the satisfiability problem for $\ABB$ interpreted 
over (either finite or infinite) interval temporal structures is EXPSPACE-complete.

\medskip
The EXPSPACE-hardness of the satisfiability problem for $\ABB$ follows from a 
reduction from the \emph{exponential-corridor tiling problem}, which is known to be 
EXPSPACE-complete \cite{convenience_of_tilings}. Formally, an instance of the 
exponential-corridor tiling problem is a tuple $\cT=(T,t_\bot,t_\top,H,$ $V,n)$ 
consisting of a finite set $T$ of tiles, a bottom tile $t_\bot\in T$, a top tile 
$t_\top\in T$, two binary relations $H,V$ over $T$ (specifying the horizontal and 
vertical constraints), and a positive natural number $n$ (represented in unary 
notation). The problem consists in deciding whether there exists a tiling 
$f:\bbN\times\{0,...,2^n-1\}\then T$ of the infinite discrete corridor of height 
$2^n$, that associates the tile $t_\bot$ (resp., $t_\top$) with the bottom (resp., 
top) row of the corridor and that respects the horizontal and vertical constraints 
$H$ and $V$, namely,
\begin{romlist}
  \item for every $x\in\bbN$, we have $f(x,0)=t_\bot$,
  \item for every $x\in\bbN$, we have $f(x,2^n-1)=t_\top$,
  \item for every $x\in\bbN$ and every $0\le y<2^n$, we have $f(x,y) \;H\; f(x+1,y)$,
  \item for every $x\in\bbN$ and every $0\le y<2^n-1$, we have $f(x,y) \;V\; f(x,y+1)$.
\end{romlist}
The proof of the following lemma, which reduces the exponential-corridor tiling problem 
to the satisfiability problem for $\ABB$, can be found in \cite{abb_report}. 
Intuitively, such a reduction exploits (i) the correspondence between the points 
$p=(x,y)$ inside the infinite corridor $\bbN\times\{0,...,2^n-1\}$ and the intervals of 
the form $I_p=[y+2^n x,y+2^n x+1]$, (ii) $\len{T}$ propositional variables which represent 
the tiling function $f$, (iii) $n$ additional propositional variables which represent (the 
binary expansion of) the $y$-coordinate of each row of the corridor, and (iv) the modal 
operators $\DA$ and $\DB$ by means of which one can enforce the local constrains over 
the tiling function $f$ (as a matter of fact, this shows that the satisfiability problem 
for the $\AB$ fragment is already hard for EXPSPACE).

\begin{lemma}\label{lemma:hardness}
There is a polynomial-time reduction from the exponential-corridor tiling problem 
to the satisfiability problem for $\ABB$.
\end{lemma}

\medskip
As for the EXPSPACE-completeness, we claim that the existence of a compass structure 
$\cG$ that features a given formula $\varphi$ can be decided by verifying suitable 
local (and stronger) consistency conditions over all pairs of contiguous rows. In fact, 
in order to check that these local conditions hold between two contiguous rows $y$ and 
$y+1$, it is sufficient to store into memory a bounded amount of information, namely, (i) 
a counter $y$ that ranges over $\bigl\{1,...,2^{2^{7\len{\varphi}}} + \len{\varphi}\cdot2^{2^{7\len{\varphi}}}\bigr\}$,
(ii) the two guessed shadings $S$ and $S'$ associated with the rows $y$ and $y+1$, and (iii) 
a function $g:S\then S'$ that captures the horizontal alignment relation between points with
an associated atom from $S$ and points with an associated atom from $S'$. This shows that 
the satisfiability problem for $\ABB$ can be decided in exponential space, as claimed by 
the following lemma. Further details about the decision procedure, including soundness and 
completeness proofs, can be found in \cite{abb_report}.

\begin{lemma}\label{lemma:completeness}
There is an EXPSPACE non-deterministic procedure that decides whether a given formula 
of $\ABB$ is satisfiable or not.
\end{lemma}

\medskip
Summing up, we obtain the following tight complexity result.

\begin{theorem}\label{th:complexity}
The satisfiability problem for $\ABB$ interpreted over (prefixes of) natural 
numbers is EXPSPACE-complete.
\end{theorem}

\section{Conclusions}\label{sec:conclusions}

In this paper, we proved that the satisfiability problem for $\ABB$ interpreted over 
(prefixes of) the natural numbers is EXPSPACE-complete. We restricted our attention to 
these domains because it is a common commitment in computer science. Moreover, this 
gave us the possibility of expressing meaningful metric constraints in a fairly natural
way. Nevertheless, we believe it possible to extend our results to the class of all
linear orderings as well as to relevant subclasses of it. Another restriction that
can be relaxed is the one about singleton intervals: all results in the paper can be 
easily generalized to include singleton intervals in the underlying structure $\bbI_N$. 
The most exciting challenge is to establish whether the modality $\bar{A}$ can be 
added to $\ABB$ preserving decidability (and complexity). It is easy to show that
there is not a straightforward way to lift the proof for $\ABB$ to 
${A\mspace{-0.3mu}B\bar{B}\bar{A}}$ (notice that $\DA$, $\DB$, and $\DBbar$ 
are all future modalities, while $\DAbar$ is a past one).

\bibliographystyle{plain}
\bibliography{biblio}


\end{document}